# SYSTEMATIC EFFECTS IN THE RADIO SOURCE PROPER MOTION

O. Titov
Geoscience Australia
PO Box 378 Canberra 2601 Australia
e-mail: oleg.titov@ga.gov.au

**Abstract**

Some models of the expanding Universe predict that the astrometric proper motion of distant radio sources embedded in space-time are non-zero as radial distance from observer to the source grows. Systematic effects due to this proper motion can even increase with distance making possible to measure them with high precision astrometric techniques like VLBI. We analyzed a large set of geodetic VLBI data spanning from 1979 till 2008 to estimate the dipole and the quadrupole harmonics in the expansion of the vector field of the proper motions of quasars in the sky. We estimated the vector spherical harmonics (three parameters for the dipole and ten - for the quadrupole systematic) directly from the VLBI group delays without intermediate calculation of the individual proper motion. The estimates have been obtained separately for different red shift zones. It was shown that the dipole harmonic does not vary significantly, whereas the amplitude of the quadrupole gradually increases with the red shift. This quadrupole pattern can be interpreted either as an anisotropic Hubble expansion, or as an indication of the primordial gravitational waves in the early Universe. However, more prosaic explanations may also be possible.

## 1 Introduction

The quasi-inertial celestial reference frame based on the accurate positions of 212 'defining' quasars is based on assumption that the physical proper motion of the radio sources are negligibly small (less than 1 $\mu$as/year) (Ma et al, 1998). However, non vanishing proper motion of distant objects in the frame of the General Relativity was predicted by Kristian and Sachs (1966). These authors showed that in the expanding Universe the proper motion of objects would increase with distance due to anisotropic expansion or the primordial gravitational waves. The corresponding systematic effect should appear in the form of second order vector spherical harmonics. The acceleration of the Solar System barycenter would cause the dipole systematic effect in the proper motion described by the first order electric type vector spherical harmonic (for example, Gwinn et al., 1997; Sovers et al., 1998; Kovalevsky, 2003; Kopeikin and Makarov, 2006). A search for this systematic started 30 years later (Pyne et al., 1996; Gwinn et al., 1997; MacMillan, 2005; Titov, 2008). The search is complicated by the presence of the intrinsic motions of the radio sources themselves. Their structural change motion can reach several hundred $\mu$as/year (Ma et al, 1998; Fey et al., 2004). Thus, the amplitude of the cosmologic effects is smaller than the random noise caused by the observational errors as well as the individual intrinsic apparent motions.

In this paper, we present the mathematical formulae used to address this problem, the most recent results obtained for the vector spherical harmonic estimates using VLBI data from 1980 to 2008, and the comparisons of the estimates made with the radio sources in different zones of red shift.

## 2 Vector spherical functions for the dipole and quadrupole systematics

Let us consider $\bar{F}(\alpha,\delta)$ as a vector field of a sphere described by the components of the proper motion vector ($\mu_\alpha \cos\delta, \mu_\delta$)

$$\bar{F}(\alpha,\delta) = \mu_\alpha \cos\delta \cdot \bar{e}_\alpha + \mu_\delta \cdot \bar{e}_\delta \tag{1}$$

where $\bar{e}_\alpha, \bar{e}_\delta$ - unit vectors. A vector field of spherical harmonics (1) can be approximated by the vector spherical functions as follows

$$\bar{F}(\alpha,\delta) = \sum_{l=1}^{\infty}\sum_{m=-l}^{l} (a_{l,m}^E \bar{Y}_{l,m}^E + a_{l,m}^M \bar{Y}_{l,m}^M) \tag{2}$$

where $\bar{Y}_{l,m}^E, \bar{Y}_{l,m}^M$ - the 'electric' and 'magnetic' transverse vector spherical functions, respectively

$$\bar{Y}_{l,m}^E = \frac{1}{\sqrt{l(l+1)}} \left( \frac{\partial V_{lm}(\alpha,\delta)}{\partial\alpha \cos\delta} \bar{e}_\alpha + \frac{\partial V_{lm}(\alpha,\delta)}{\partial\delta} \bar{e}_\delta \right)$$

$$\bar{Y}_{l,m}^M = \frac{1}{\sqrt{l(l+1)}} \left( \frac{\partial V_{lm}(\alpha,\delta)}{\partial\delta} \bar{e}_\alpha - \frac{\partial V_{lm}(\alpha,\delta)}{\partial\alpha \cos\delta} \bar{e}_\delta \right)$$

The function $V_{l,m}(\alpha,\delta)$ is given by

$$V_{l,m}(\alpha,\delta) = (-1)^m \sqrt{\frac{(2l+1)(l-m)!}{4\pi(l+m)!}} P_l^m(\sin\delta) e^{im\alpha}$$

where $P_l^m(\sin\delta)$ - the associated Legendre functions.
The coefficients of expansion (2) $a_{l,m}^E, a_{l,m}^M$ to be estimated as follows

$$a_{l,m}^E = \int_0^{2\pi}\int_{-\pi/2}^{\pi/2} \bar{F}(\alpha,\delta) \bar{Y}_{l,m}^E {}^*(\alpha,\delta) \cos\delta\, d\alpha\, d\delta$$

$$a_{l,m}^M = \int_0^{2\pi}\int_{-\pi/2}^{\pi/2} \bar{F}(\alpha,\delta) \bar{Y}_{l,m}^M {}^*(\alpha,\delta) \cos\delta\, d\alpha\, d\delta$$

where * means a complex conjugation. This system of equations can be solved by the least squares method. In this research the coefficients are estimated as global parameters from a large set of VLBI data.

The dipole systematic effect (secular aberration drift) in proper motion ($\mu_\alpha \cos\delta, \mu_\delta$) (1) is given by the electric-type vector spherical harmonics of the first order (Gwinn et al., 1997; Kopeikin and Makarov, 2006}

$$\mu_\alpha \cos\delta = \frac{1}{c}(-a^E_{1,1}\sin\alpha + a^E_{1,-1}\cos\alpha)$$
$$\mu_\delta = \frac{1}{c}(-a^E_{1,1}\sin\delta\cos\alpha - a^E_{1,-1}\sin\delta\sin\alpha + a^E_{1,0}\cos\delta) \tag{3}$$

where $c$ - the speed of light, $a^E_{1,0}, a^E_{1,1}, a^E_{1,-1}$ - components of the vector of the secular aberration drift.

The proper motion introduced by the electric- and magnetic-type vector spherical harmonics of the second order are given by

$$\bar\mu(\alpha,\delta) = \sum_{m=-2}^{2}(a^E_{2,m}\bar Y^E_{2,m} + a^M_{2,m}\bar Y^M_{2,m}) \tag{4}$$

For instance, anisotropic expansion of the Universe makes two coefficients to be non-zero. The Hubble law for isotropic Universe $V = HR$ links the recession velocity $V$ and distance $R$. In the case of anisotropic expansion for the it should be replaced by

$$V = (e_{33}\sin^2\delta + \frac{1}{2}(e_{11}+e_{22})\cos^2\delta + \frac{1}{2}(e_{11}-e_{22})\cos 2\alpha\cos^2\delta)R =$$
$$= (H + \Delta H_3 \sin^2\delta + \frac{\Delta H_{12}}{2}\cos 2\alpha \cos^2\delta)R$$

where the coefficients $e_{11}, e_{22}, e_{33}$ serve as the diagonal elements of the expansion tensor, $H = \frac{1}{2}(e_{11}+e_{22})$ is the Hubble constant and two parameters that describe the Hubble constant anisotropy are given by

$$\Delta H_3 = e_{33} - \frac{1}{2}(e_{11}+e_{22})$$
$$\Delta H_{12} = e_{11} - e_{22} \tag{5}$$

The transversal proper motion due to the anisotropic expansion is are given by

$$\mu_\alpha \cos\delta = -\frac{1}{2}(e_{11} - e_{22})\sin 2\alpha \cos\delta = -\frac{\Delta H_{12}}{2}\sin 2\alpha \cos\delta$$

$$\mu_\delta = (e_{33} - \frac{1}{2}(e_{11} + e_{22}))\cos\delta \sin\delta - \frac{1}{2}(e_{11} - e_{22})\cos 2\alpha \sin\delta \cos\delta = \qquad (6)$$

$$= \Delta H_3 \sin\delta \cos\delta - \frac{\Delta H_{12}}{2}\cos 2\alpha \sin\delta \cos\delta$$

It can be shown from (4) and (6) that $a_{2,0}^E = \frac{\Delta H_3}{2}$; $a_{2,2}^E = \frac{\Delta H_{12}}{4}$

In the case of shear-free isotropic expansion of the Universe, the Hubble constant is uniform around the sky. The diagonal elements now $e_{11} = e_{22} = e_{33}$, then $\Delta H_3 = \Delta H_{12} = a_{2,0}^E = a_{2,2}^E = 0$ and the transverse proper motion vanishes.

The effect of gravitational waves in the radio source proper motion could be detected as the both electric- or magnetic-type vector spherical harmonics. Pyne et al. (1996), Gwinn et al. (1997) assumed that the wavelength of gravitational waves is less than the Hubble length. Under this assumption, the amplitudes of the second order harmonics should be similar for the radio sources across all range of distance. Kristian and Sachs (1966) considered arbitrary gravitational waves, and the proper motion, calculated on their approach, are proportional to the distance to a radio source, and, consequently to the red shift.

## 3 Data and modelling

The first and second degree spherical harmonics were estimated by the least squares collocation method Titov (2004). The database comprises of about 4.3 million observations of group delay with different baseline and sources made in 3724 24-hour sessions between April, 1980 and September, 2008. The equatorial coordinates of more than 2000 radio sources were observed as global or 'arc' parameters (see solution description below). The Earth orientation parameters, nutation offsets correction to the IAU2000 model as well as station coordinates were estimated as 'arc' parameters. No-net-rotation (NNR) and no-net-translation (NNT) constraints imposed the station positions for each 24-hour session. Clock offsets, troposphere wet delays and north-south and east-west gradients were estimated as stochastic parameters for each observational epoch. The vector spherical harmonics were treated as global parameters, similar to the approached used by MacMillan (2005}.

## 4 Results and comparison

In the 'basic' solution, all radio sources observed in geodetic and astrometric VLBI programs were considered as global parameter, except for the 102 'other' radio sources (Ma et al., 1998), treated as local or 'arc' parameters. In other solutions we considered as global only the radio sources inside one of five zones of red shift as listed in Table 1.

Estimates of the secular aberration drift components (3) are presented in Table 1. The amplitude of the effect varies in the range of 15-24 μas/year with 1σ standard error of 1-2 μas/year for different red shift zones. Figure 1 shows the dipole systematic for the 'basic' solution.

The coordinates of the vector of the secular aberration drift is stable for right ascension for all solutions. Declination gradually changes from δ = 35° +/- 6° for the 'close' radio sources to δ = -9° +/- 11° for the 'distant' radio sources. The standard error for the latter case is larger due to fewer radio sources at high red shift.

We also presented another set of solutions (Table 2). The second order vector spherical harmonics (6) were estimated along with the dipole ones (3). The sums of systematic (dipole and quadrupole) effects in proper motion for the 'basic' solution are shown on Figure 2.

The estimates of the quadrupole effect amplitude are statistically significant (standard error is about 15-20% except for the first zone), in spite of the limited number of the reference radio sources within each red shift zone. For the most distant sources (mean $z$ = 2.6) the amplitude is unexpectedly large (58 +/- 10 $\mu$as/year), as shown on Figure 3. The dipole effect parameters in each zone of red shift are only slightly changed with respect to the similar solutions from Table 1 (within ~ 5 $\mu$as/year in absolute values). The amplitude of the dipole effect is stable along all these zones again, whereas, the quadrupole systematic gradually increases with red shift (Figure 4). It can be concluded that the quadrupole effect might be caused by primordial gravitational waves in accordance with the early prediction by Kristian and Sachs (1966).

However, for narrow red shift zones the number of radio sources is very limited, especially at high red shift. It resulted in large correlations between first and second order vector spherical harmonics (correlation coefficients up to 0.8-0.9). More observations of distant radio sources in the southern hemisphere (under δ = -40°) should be undertaken in order to make more reliable conclusion.

| Range of red shift | | 0<z<0.7 | 0.5<z<1.5 | All sources | 1.0<z<3.0 | 1.5<z<3.0 | z>1.7 |
|---|---|---|---|---|---|---|---|
| Number of sources | | 357 | 468 | 1530 | 542 | 312 | 287 |
| mean z | | 0.44 | 0.95 | 1.31 | 1.86 | 2.23 | 2.61 |
| Electric-type harmonics ($\mu$as/year) | $a_{1,0}^E$ | 13.3 +/- 2.1 | 7.6 +/- 2.0 | 7.9 +/- 1.7 | 2.7 +/- 2.0 | 0.9 +/- 2.4 | -2.5 +/- 2.8 |
| | $a_{1,1}^E$ | 0.1 +/- 1.1 | 3.6 +/- 1.1 | -1.2 +/- 0.9 | 0.3 +/- 1.1 | 0.7 +/- 1.5 | 10.8 +/- 2.1 |
| | $a_{1,-1}^E$ | -18.9 +/- 1.0 | -16.9 +/- 1.0 | -20.3 +/- 0.8 | -15.4 +/- 1.2 | -15.5 +/- 1.6 | -10.7 +/- 2.0 |
| Dipole amp ($\mu$as/year) | | 23.5 +/- 1.5 | 18.9 +/- 1.2 | 21.8 +/- 1.0 | 15.9 +/- 1.2 | 15.5 +/- 1.6 | 15.4 +/- 2.0 |
| α (deg) | | 270.6 +/- 3.3 | 287.8 +/- 4.2 | 266.6 +/- 2.6 | 280.3 +/- 4.8 | 272.6 +/- 5.8 | 315.3 +/- 10.7 |
| δ (deg) | | 35.1 +/- 5.7 | 23.7 +/- 6.9 | 21.2 +/- 5.0 | 9.7 +/- 7.8 | 3.4 +/- 9.0 | -9.2 +/- 11.9 |

Table 1. Estimates of the vector spherical harmonics $l$=1 for different sets of the reference radio sources, where $z$ – red shift, (α, δ) – positions of the vector of secular aberration drift for different sets of reference radio sources

| Range of red shift | | 0<z<0.7 | 0.5<z<1.5 | All sources | 1.0<z<3.0 | 1.5<z<3.0 | z>1.7 |
|---|---|---|---|---|---|---|---|
| Number of sources | | 357 | 468 | 1530 | 542 | 312 | 287 |
| mean z | | 0.44 | 0.95 | 1.31 | 1.86 | 2.23 | 2.61 |
| Electric-type harmonics ($\mu$as/year) | $a_{1,0}^{E}$ | 12.1 +/- 3.1 | 0.4 +/- 3.0 | 6.1 +/- 3.0 | -0.4 +/- 3.2 | 3.7 +/- 3.9 | -0.4 +/- 5.5 |
| | $a_{1,1}^{E}$ | -2.5 +/- 1.6 | -0.1 +/- 1.4 | 0.0 +/- 1.2 | -9.2 +/- 1.7 | -7.3 +/- 3.0 | 7.9 +/- 4.2 |
| | $a_{1,-1}^{E}$ | -17.7 +/- 1.3 | -12.6 +/- 1.4 | -25.6 +/- 1.1 | -16.7 +/- 2.0 | -17.0 +/- 3.2 | -18.4 +/- 3.9 |
| Dipole amp ($\mu$as/year) | | 21.5 +/- 2.1 | 12.6 +/- 1.4 | 26.3 +/- 1.3 | 19.1 +/- 2.0 | 18.9 +/- 3.2 | 20.0 +/- 3.8 |
| α (deg) | | 270.6 +/- 3.3 | 269.5 +/- 6.3 | 270.0 +/- 2.7 | 241.2 +/- 7.2 | 246.7 +/- 12.4 | 293.3 +/- 15.3 |
| δ (deg) | | 35.1 +/- 5.7 | 1.6 +/- 13.9 | 13.4 +/- 6.9 | -1.2 +/- 9.9 | 3.4 +/- 14.0 | -1.0 +/- 16.0 |
| Electric-type harmonics ($\mu$as/year) | $a_{2,0}^{E}$ | 1.2 +/- 1.9 | 7.1 +/- 1.7 | 2.5 +/- 1.5 | 2.9 +/- 1.9 | -3.1 +/- 2.6 | 0.1 +/- 4.5 |
| | $a_{2,1}^{E}$ | 1.4 +/- 1.0 | -1.7 +/- 1.0 | 5.2 +/- 0.7 | 2.0 +/- 1.7 | -7.2 +/- 3.2 | -5.2 +/- 3.4 |
| | $a_{2,-1}^{E}$ | -0.4 +/- 1.3 | 6.2 +/- 1.1 | -1.1 +/- 0.8 | 9.5 +/- 1.4 | -7.1 +/- 2.7 | 0.2 +/- 3.6 |
| | $a_{2,2}^{E}$ | -3.9 +/- 0.8 | 2.3 +/- 0.7 | 1.9 +/- 0.6 | 7.1 +/- 0.9 | 10.9 +/- 1.9 | 42.6 +/- 3.3 |
| | $a_{2,-2}^{E}$ | -3.8 +/- 0.8 | -4.0 +/- 0.7 | -2.8 +/- 0.6 | -1.0 +/- 0.9 | 0.7 +/- 1.9 | -5.6 +/- 2.6 |
| Magnetic-type harmonics ($\mu$as/year) | $a_{2,0}^{M}$ | -5.9 +/- 1.0 | -7.6 +/- 0.8 | -7.3 +/- 0.7 | -12.4 +/- 1.1 | -13.8 +/- 1.7 | 3.6 +/- 3.0 |
| | $a_{2,1}^{M}$ | -3.8 +/- 0.9 | -3.7 +/- 0.8 | 0.3 +/- 0.5 | -14.3 +/- 1.4 | -1.8 +/- 2.8 | -1.5 +/- 2.8 |
| | $a_{2,-1}^{M}$ | -1.3 +/- 0.9 | -8.2 +/- 0.8 | -2.0 +/- 0.5 | -4.5 +/- 1.3 | 2.9 +/- 2.9 | 3.4 +/- 2.9 |
| | $a_{2,2}^{M}$ | 3.7 +/- 0.8 | 3.3 +/- 0.7 | -0.5 +/- 0.6 | 1.6 +/- 0.9 | 2.8 +/- 2.1 | 14.2 +/- 3.4 |
| | $a_{2,-2}^{M}$ | -1.7 +/- 0.9 | -1.1 +/- 0.7 | 2.4 +/- 0.5 | 2.6 +/- 0.9 | 12.7 +/- 1.7 | 36.0 +/- 2.5 |
| Quadrupole amp ($\mu$as/year) | | 10.1 +/- 3.4 | 16.2 +/- 3.0 | 10.5 +/- 2.4 | 23.2 +/- 4.1 | 24.5 +/- 7.6 | 58.3 +/- 10.3 |

Table 1. Estimates of the vector spherical harmonics $l=1,2$ for different sets of the reference radio sources, where $z$ – red shift, (α, δ) – positions of the vector of secular aberration drift for different sets of reference radio sources

**Acknowledgments**

I thank David Jauncey for interesting discussion and helpful comments on this paper.

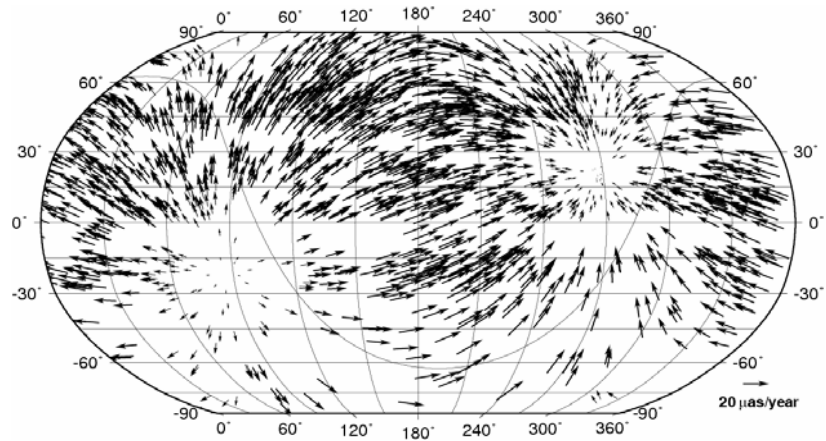

Figure 1. Dipole systematic for the 'basic' solution ('all sources' in Table 1)

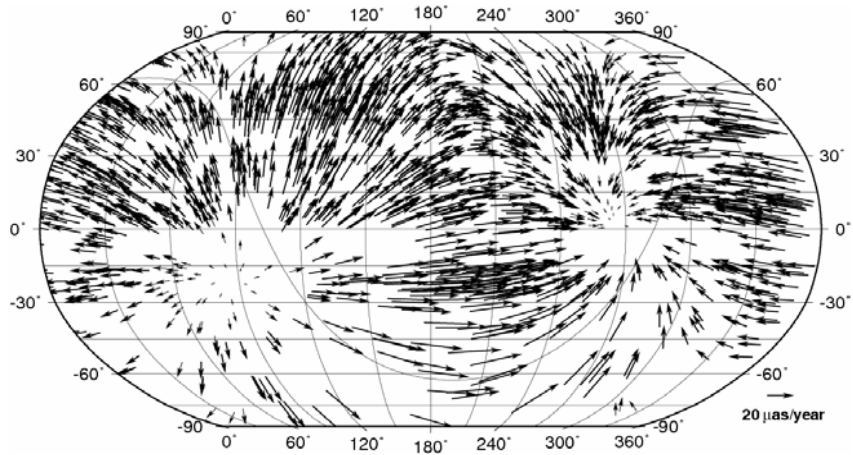

Figure 2. Dipole and quadrupole systematic for the 'basic' solution ('all sources' in Table 2)

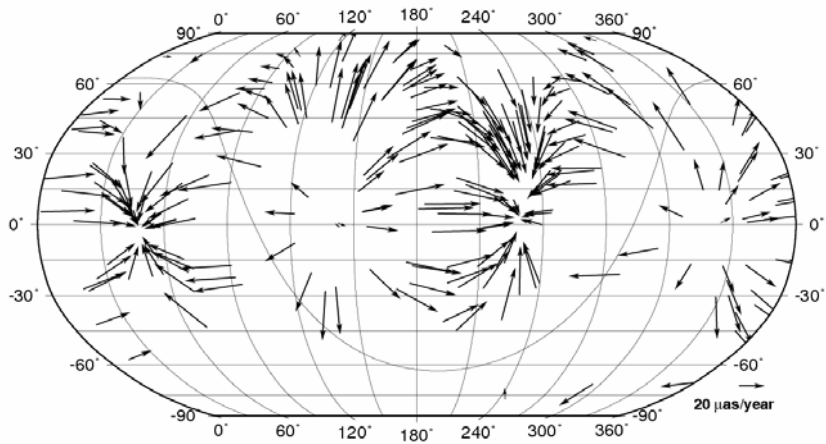

Figure 3. Quadrupole systematic for the most distant radio sources (last column in Table 2)

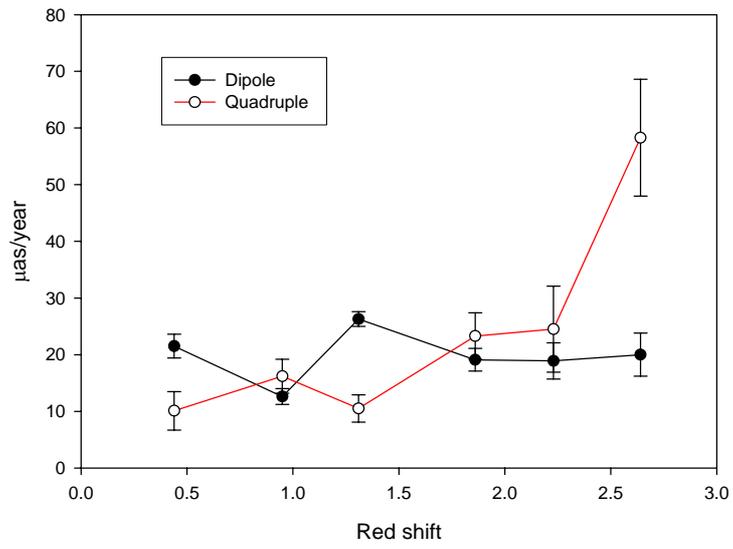

Figure 4. Amplitude of the dipole and quadrupole systematic with respect of the mean red shift